\documentclass[twocolumn]{aastex6}
\usepackage[]{natbib}
\usepackage{graphicx, amsmath}
\usepackage{array}
\usepackage{booktabs}
\usepackage{float}
\usepackage[caption=false]{subfig}
\usepackage{color}

\begin{document}

\title{Photometric Detection of Multiple Populations in Globular Clusters Using Integrated Light}
\shorttitle{Multiple Populations in Globular Clusters}
\author{William P. Bowman\altaffilmark{1}, Catherine A. Pilachowski, Liese van Zee, Amanda Winans}
\affil{Department of Astronomy, Indiana University, Bloomington, IN 47405}
\email{bowman@psu.edu}
%, cpilacho@indiana.edu, lvanzee@indiana.edu}

\and

\author{Robin Ciardullo\altaffilmark{1}, Caryl Gronwall\altaffilmark{1}}
\affil{Department of Astronomy \& Astrophysics, The Pennsylvania
State University, University Park, PA 16802}
%\email{rbc@astro.psu.edu, caryl@astro.psu.edu}

\altaffiltext{1}{Institute for Gravitation and the Cosmos, The Pennsylvania
State University, University Park, PA 16802}

\begin{abstract}
We investigate the multiple stellar populations of the globular clusters
M3, M5, M13 and M71 using $g^\prime$ and intermediate-band CN-$\lambda 3883$ photometry 
obtained with the WIYN 0.9-m telescope on Kitt Peak. We find a strong correlation between red giant
stars' CN$-g^\prime$ colors and their spectroscopic sodium abundances, thus demonstrating
the efficacy of the two-filter system for stellar population studies.  In all four clusters, 
the observed spread in red giant branch CN$-g^\prime$ colors is wider than that expected 
from photometric uncertainty, confirming the well-known chemical inhomogeneity of these systems.
M3 and M13 show clear evidence for a radial dependence in the CN-band strengths of its red giants, 
while the evidence for such a radial dependence of CN strengths in M5 is ambiguous. 
Our data suggest that the dynamically old, relatively metal-rich M71 system is well mixed, as it shows 
no evidence for chemical segregation.  Finally, we measure the radial gradients in the
integrated CN$-g^\prime$ color of the clusters and find that such gradients are easily detectable 
in the integrated light.  We suggest that photometric observations of color gradients within globular 
clusters throughout the Local Group can be used to characterize their multiple populations, 
and thereby constrain the formation history of globular clusters in different galactic environments.
\end{abstract}


\section{Introduction}

For most of the twentieth century, Milky Way globular clusters (GCs) were thought of as simple stellar 
populations, defined by a single age and metallicity.  However, in the past several years, 
our view of these systems has changed, and it is now widely accepted that globular clusters contain
multiple stellar populations \citep[e.g.,][]{bedin2004, gratton2012, carretta2010, villanova2007, milone2012, 
piotto2015}.  Spectroscopic analyses have found that significant variations in the abundances of light 
elements, especially those associated with proton-capture processes, such as C, N, O, F, Na, Mg, and 
Al \citep[see][and references therein]{gratton2012} exist within most clusters.   The physical 
mechanism responsible for these differences is still uncertain, with possible sources
being the CNO-processed winds of massive, fast-rotating stars formed early in the
history of the cluster and the superwinds of intermediate-mass asymptotic giant branch (AGB) stars
\citep{gratton2012}.  Regardless of the cause, the dispersions in the abundances of these light elements,
which have been measured in a number of globular clusters, are thought to 
indicate that globular clusters are more than simply single-age, single-metallicity stellar systems.

The most secure way of detecting abundance variations in the stars of globular clusters is through careful 
spectroscopic analyses.  However, the presence of multiple stellar populations within a 
globular cluster can also be revealed photometrically by examining features in the 
color-magnitude diagram (CMD), such as bifurcations in the main sequence \citep[e.g.,][]{milone2015} 
and red giant branch \citep[e.g.,][]{kayser2008, milone2016}.   Numerous investigations of this type 
have been performed in the past few years \citep[e.g.,][]{hsyu2014, roh2011, lee2013, milone2017},
though, with a few notable exceptions  \citep[e.g.,][]{cummings2014, lee2015}, the filter sets used were 
not optimized for abundance studies.   This is unfortunate, since, as \citet{sbordone2011} have 
demonstrated, filter systems such as those defined by \citet{stromgren1966} can greatly increase the sensitivity
of photometry to chemical inhomogeneities.    Such observations open up the possibility of studying 
abundance dispersions within the globular clusters of other galaxies, thus allowing studies of the physics of
GC formation under conditions far different from those that existed in the Milky Way.

Spectroscopic studies of nearby globular clusters have found that sodium abundance correlates with 
stellar population, with second generation stars having higher abundances \citep{carretta2010}.  Since 
sodium often correlates with carbon and nitrogen \citep{gratton2012}, one should be able to use
CN-band strengths as a tracer of stellar population.  For example, \citet{smith2009}, \citet{smith2013},
\citet{smith2015a}, and \citet{smith2015b} have used 
literature measurements of spectroscopically determined CN, O, and Na line strengths
to demonstrate a clear correlation between CN and [Na/Fe] in the globular clusters
NGC 288, NGC 362, M5, 47 Tucanae, and M71.
A clear correlation between CN-band strength and [Na/Fe] was found 
in each of these clusters.
These studies also found significant variations in the CN-band
strength within the clusters, consistent with the results of \citet{smith2006} and \citet{martell2009}.

Here, we examine the distribution of stellar populations within intermediate-metallicity globular clusters 
using narrow band photometry centered on the CN absorption complex at $\lambda \sim 3883$~\AA.
In \S2 and \S3, we describe our photometric observations of four well known Galactic globular clusters,
and detail the reduction techniques used to measure CN$-g^\prime$ colors for the systems' red giant
branch (RGB) stars.  In \S4 we discuss the correlations between CN$-g^\prime$ colors and [Na/Fe] abundance, and
in \S5,  we characterize the photometric signatures of these abundance variations.  In \S6, we examine
the behavior of the CN$-g^\prime$ colors of globular cluster integrated light, and show that stochastic
effects associated with stellar evolution are generally unimportant for such a measurement.  We conclude
by summarizing our results, and their applicability to studies of multiple
populations in globular clusters in nearby galaxies.

\section{Target Selection}

We chose as the targets for our study four nearby globular clusters that have a host of
spectroscopic abundance measurements.  Three of the clusters, M3, M5, and M13, are 
dynamically young ($t(r_{h}) \geq 2$~Gyr), bright
($M_V < -8.8$), intermediate-metallicity ([Fe/H] $\lesssim -1.3$) systems with little foreground reddening
($E(B-V) \lesssim 0.03$); the fourth (M71), is a relatively faint ($M_V \sim -5.6$), nearby
cluster ($d \sim 4$~kpc), that has a shorter median relaxation time ($t(r_{h}) = 0.27$~Gyr),
higher metallicity ([Fe/H] $\sim -0.8$) and greater foreground extinction 
($E(B-V) = 0.25$); see Table~\ref{tab:gc}.
All four systems have mass-to-light ratios ($M/L_{V} \leq 2.5$) that 
confirm their identity as true globular clusters, rather than accreted dwarf galaxies
\citep{kimmig2015, kamann2014}.  Table~\ref{tab:gc} summarizes some of the relevant cluster
parameters, along with the $u^\prime - g^\prime$ color of the main-sequence turnoff, as derived from
Sloan Digital Sky Survey (SDSS) photometry \citep{an2008}.  We use the latter values to establish the
zero point of the instrumental magnitudes.

\floattable
\begin{deluxetable}{ccccccccc}
\tablecaption{Cluster parameters \label{tab:gc}}
%\tablecolumns{7}
%\tablenum{1}
\tablewidth{0pt}
\tablehead{
&&\colhead{Distance\tablenotemark{a}} 
&\colhead{Mass\tablenotemark{b}} &&&\colhead{${\rm t(r_{h})}$\tablenotemark{c}} 
&\colhead{Half-light Radius, ${\rm r_{e}}$\tablenotemark{a}}  &\colhead{Turnoff\tablenotemark{d}}\\
\colhead{Cluster} &\colhead{$M_V$\tablenotemark{a}} &\colhead{(kpc)} &\colhead{($10^5 \, M_{\odot}$)}  
&\colhead{[Fe/H]\tablenotemark{a}} & \colhead{$E(B-V)$\tablenotemark{a}} 
& \colhead{(Gyr)} &\colhead{(arcmin)} &\colhead{$u^\prime-g^\prime$}
}
\startdata
 M3   &$-8.9$ &10.2 & 5.7 &$-1.50$ & 0.01 & 6.2   & 2.31  & 1.0  \\
 M5   &$-8.8$ &7.5   & 5.7 &$-1.29$ & 0.03 & 2.6   & 1.77  & 1.0  \\ 
 M13 &$-8.5$ &7.1   & 5.6 &$-1.53$ & 0.02 & 2.0   & 1.69  & 0.7  \\ 
 M71 &$-5.6$ &4.0   & 4.4 &$-0.78$ & 0.25 & 0.27 & 1.67  & 1.5  \\ 
\enddata
\tablenotetext{a}{From \citet{harris2010}. }
\tablenotetext{b}{From \citet{vandenberg2013}. }
\tablenotetext{c}{From \citet{kimmig2015} and \citet{kamann2014}. }
\tablenotetext{d}{From \citet{an2008}. }
\end{deluxetable}

\section{Observations and Data Processing}

Images of the four program globular clusters were obtained using the Half Degree Imager (HDI) 
on the f/7.5 WIYN 0.9-m telescope at Kitt Peak with both a Sloan $g^\prime$ filter and a 70~\AA -wide 
interference filter centered on the CN absorption trough at 3883~\AA. The image scale 
of the data is $0\farcs 43$~per pixel and the images typically have a full width at half maximum
of $\sim 3.5$~pixels.  The observations are summarized in Table~\ref{tab:obs}.

\floattable
\begin{deluxetable}{cccccc}
\tablecaption{Table of Observations \label{tab:obs}}
%\tablecolumns{5}
%\tablenum{2}
\tablewidth{0pt}
\tablehead{
& &\colhead{Number of} &\colhead{Exp.~Time} && \\
\colhead{Cluster} &\colhead{Filter} &\colhead{Exposures} & \colhead{(sec)} 
&\colhead{Seeing} & \colhead{Date} }
\startdata
 M3 & CN             & 3  & 900 &$1\farcs 5$  & 7-Jul-2015 \\
 M3 & $g^\prime$  & 5  & 30   &$1\farcs 2$  & 4-Jul-2015 \\
 M5 & CN             & 3  & 900 &$1\farcs 4$  & 7-Jul-2015 \\
 M5 & $g^\prime$  &10 & 30   &$2\farcs 1$  & 6-Jul-2015 \\
 M13 & CN           & 4  & 900 &$1\farcs 7$  & 6-Jul-2015 \\
 M13 &$g^\prime$ & 5  & 30   &$1\farcs 7$  & 4-Jul-2015 \\
 M71 & CN           & 3  & 900 &$1\farcs 6$  & 7-Jul-2015 \\
 M71 &$g^\prime$ & 5  & 30   &$1\farcs 3$  & 7-Jul-2015 \\ 
\enddata
%\tablenotetext{a}{}
%\tablecomments{}
\end{deluxetable}

The data were processed using standard IRAF\footnote{IRAF is distributed by the National Optical 
Astronomy Observatory, which is operated by the Association of Universities for Research in Astronomy
(AURA) under a cooperative agreement with the National Science Foundation.} photometric CCD
reduction techniques.  After bias subtraction, flatfielding, and cosmic-ray rejection, the frames of each
cluster were median-combined to create a high signal-to-noise pair of CN and $g^\prime$ images.
An astrometric solution to each field was then obtained using the coordinates of reference stars in the
USNO-B1.0 catalog \citep{monet2003}, and crowded field 
PSF-fitting photometry was performed using IRAF/DAOPHOT \citep{stetson1987}, following the 
procedures outlined by \citet{davis1994}. Finally, to place the $g^\prime$ data on an absolute scale, we used 
the astrometric coordinates of each star to identify its counterpart in the SDSS catalog \citep{an2008}.
After excluding known variable stars \citep{samus2009} and objects within $\sim 1$ effective radius 
of the cluster center (which have poor SDSS photometry due to crowding), we computed the magnitude 
difference between our instrumental $g^\prime$ magnitudes and those found by SDSS.
For red giants measured from our shallower images, we found that crowding may contribute to increased 
photometric errors within 20\arcsec\ of the cluster center, and we therefore excluded stars within that 
radius in M3, M5, and M13 from our analysis. 
M71 is sufficiently sparse that it requires no such exclusion of central stars.

\begin{figure*}
\centering
\noindent\includegraphics[width=0.8\linewidth]{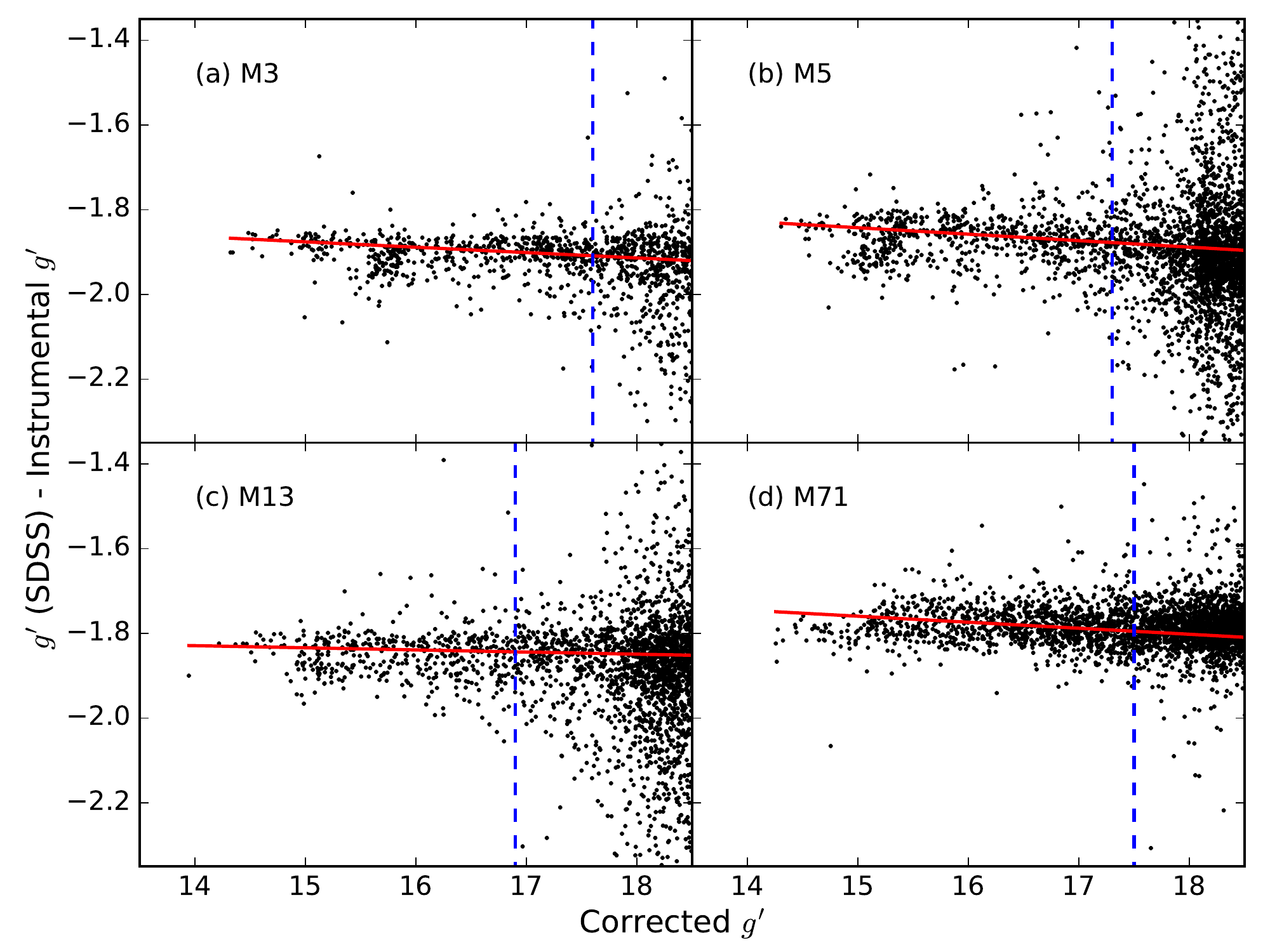}
\caption{Our instrumental $g^\prime$ magnitudes compared to those recorded in the
  SDSS catalog \citep{an2008}.  Variable stars and blended objects have been removed from the 
  sample.   The best-fit regression lines are shown in red; the shallow slope of these lines is due to a 
  color term, and is unimportant for our analysis since any residual slope is removed when we 
  fit a fiducial red giant branch.   The dotted vertical lines show the faint limit 
  of the red giant branch.     The scatter at the bright end suggests that
  the one sigma photometric errors for RGB stars is $\sim 3\%$. }
\label{fig:instrumental}
\end{figure*}

Figure~\ref{fig:instrumental} displays the residuals between our measurements and that of the SDSS catalog.   
The slight slope in the stellar locii, $\sim 0.01~\Delta g^\prime/g^\prime_{\rm inst}$,
is due to the presence of a small color-term between the SDSS and WIYN $g^\prime$ systems
and is of no significance for our analysis, as it is removed when we 
fit a fiducial red giant branch.   Note that in the magnitude range of our 
program's red giant branch stars (the faint end is identified by the dotted lines in the figure), 
the scatter between our $g^\prime$ 
measurements and those of SDSS is $\sim 4\%$, independent of stellar luminosity.  Because the 
photometric error associated with SDSS $g^\prime$ photometry is $\sim 2\%$, this suggests that 
the internal uncertainty associated with our $g^\prime$ magnitudes is $\sim 3\%$.    This number is
considerably larger than the internal errors reported by the DAOPHOT {\tt allstar} algorithm.

\begin{figure*}
\centering
\noindent\includegraphics[width=0.794\linewidth]{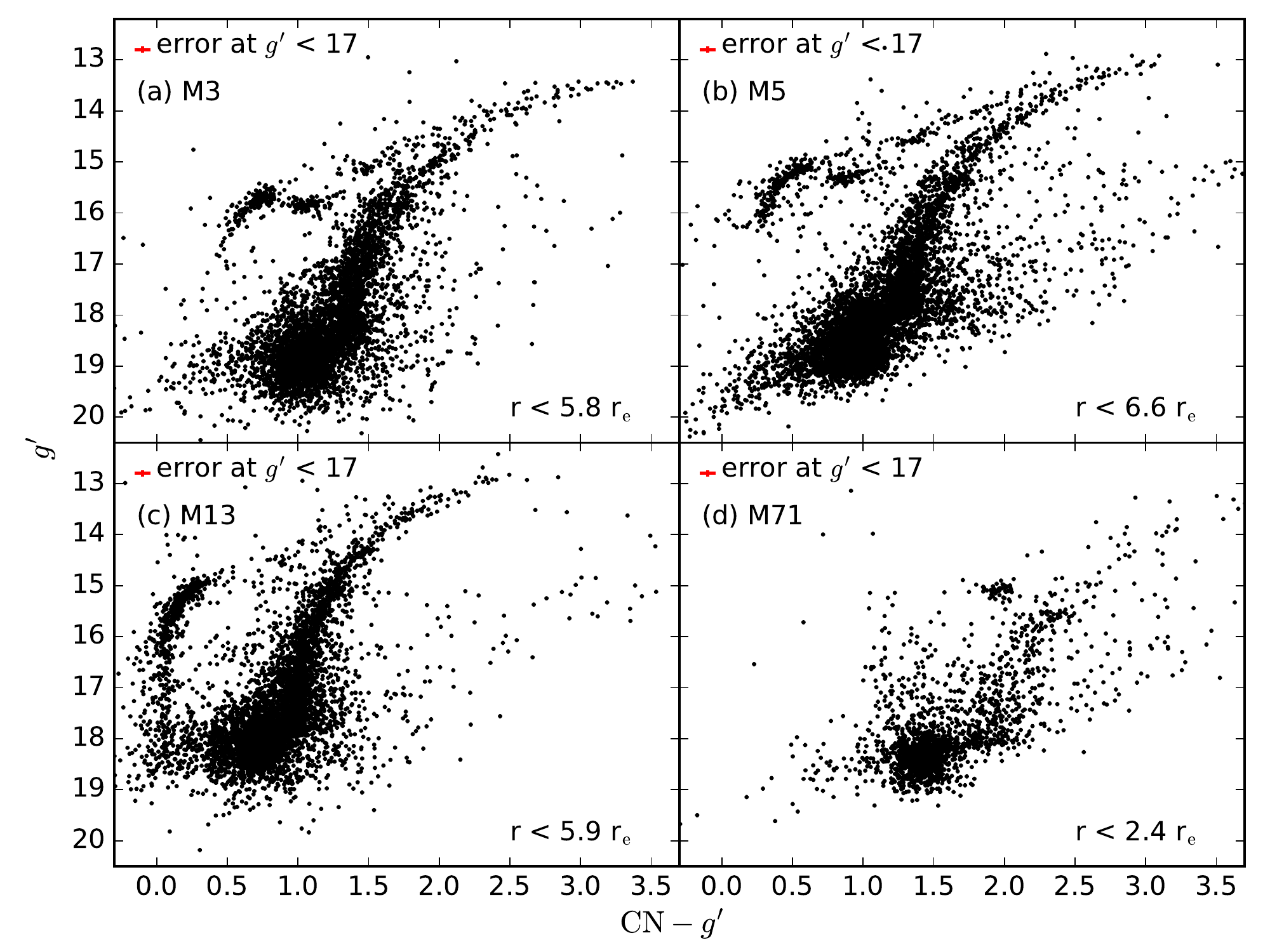}
\caption{Color-magnitude diagrams for the four GCs, with $g^\prime$ magnitudes plotted against 
CN$-g^\prime$ colors. The CN zero point is based on the SDSS $u^\prime$ magnitude at the main-sequence turnoff.
Variable stars have been removed from the figure.  The typical one sigma photometric
error of giant branch stars is shown via the red cross in the upper right hand corner of
each panel.  Note that the width of the giant branch is much larger than that predicted from the 
photometric errors.}
\label{fig:cmd}
\end{figure*}

The zero point of our CN-band photometry was set to match that of the
SDSS $u^\prime$ system.  Specifically, we took the SDSS $u^\prime - g^\prime$ color of each globular 
cluster's main-sequence turnoff  (given in Table~\ref{tab:gc}) and shifted our CN magnitudes so that the 
CN$-g^\prime$ color of the cluster's main-sequence turnoff matched this color. Figure~\ref{fig:cmd} 
displays the resultant color-magnitude diagrams (CMDs). Note that because our analysis utilizes 
differential colors relative to a fiducial giant branch, this zero point is arbitrary.

\section{Spectroscopic versus Photometric Abundances}

The goal of our experiment is to measure the sensitivity of RGB CN$-g\prime$ colors to
changes in the stellar population.  This can most easily be done by using red giant branch
stars whose chemical abundances have been determined spectroscopically.  In particular, the
proton-capture element sodium, when normalized to iron, is an excellent indicator of chemical
enrichment due to proton-capture nucleosynthesis.

Five samples of [Na/Fe] abundances are suitable for our analysis.   Our primary source of
abundance data is the WIYN-Hydra based measurements of \citet{johnson2005}, who determined
[Na/Fe] values for 76 stars in M3 and 112 stars in M13 to a precision of 0.05~dex.   For the 
globular cluster M71, the [Na/Fe] measurements come from two sources:  the 33 WIYN-Hydra 
observations of \citet{cordero2014}, which have an abundance precision of 0.05~dex, and the 25 
stars measured by \citet{ramirez2002} using the Keck Telescope's High Resolution Echelle 
Spectrograph (0.08~dex precision).   An abundance comparison of the six stars common to both 
studies indicates the two samples are statistically consistent and their error estimates are reasonable.   
The [Na/Fe] measurements for the cluster M5 are also taken from two
sources. \citet{cordero2014} measured [Na/Fe] values for 61 stars in M5, each with a precision of 0.05~dex,
and \citet{carretta2009a} and \citet{carretta2009b} measured 122 stars using the ESO UVES and GIRAFFE 
fiber-fed spectrographs with typical uncertainties of 0.05~dex. The abundance measurements for 
16 stars that appear in both samples are in good agreement.
All [Na/Fe] abundances assume local thermal equilibrium (LTE) conditions.

In total, the sample of stars with known [Na/Fe] abundances consists of 407 objects.  The largest
[Na/Fe] spread is displayed by the stars in M13, where $\Delta$[Na/Fe] is 1.37~dex,
while the smallest spread is in M71, where $\Delta$[Na/Fe] = 0.92~dex.  M3 and M5 are
intermediate cases, with $\Delta$[Na/Fe] = 1.00~dex. 
\citet{carretta2010} found that the extent of the Na-O anti-correlation, as well as the range 
of the [Na/Fe] abundances, among stars in a cluster correlate with cluster mass, 
metallicity, and Galactic origin (inner/outer halo vs. disk).

\begin{figure*}
\centering
\noindent\includegraphics[width=0.8\linewidth]{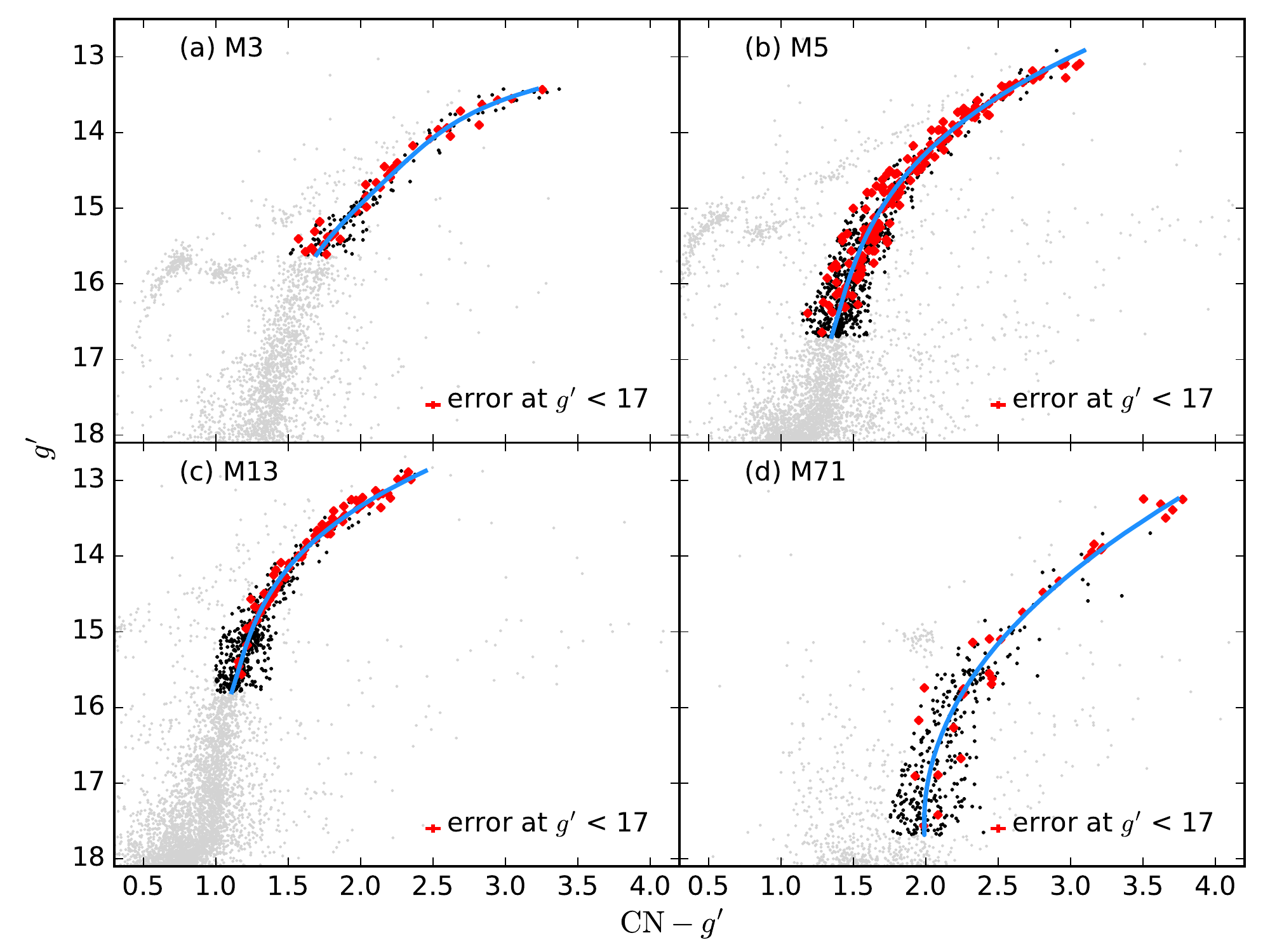}
\caption{The red giant branch of the CMD of each cluster.  Our fiducial mean
RGB track is displayed in blue.  The stars that were used to produce this track are
shown in black, while stars not used in the fit are displayed in grey.   Stars with known 
[Na/Fe] abundances are identified with red diamonds.}
\label{fig:rgb-cmd}
\end{figure*}

\begin{figure*} %[h!]
\centering
\noindent\includegraphics[width=0.9\linewidth]{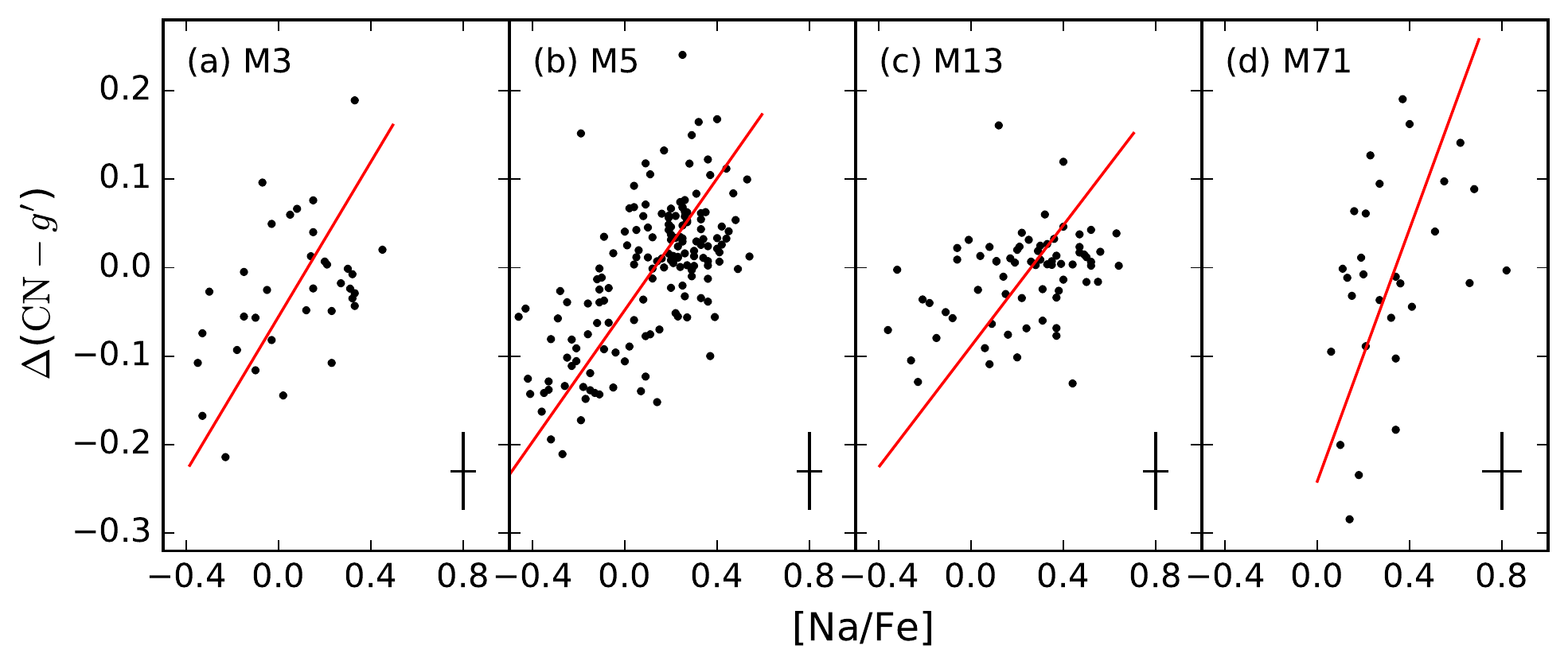}
\caption{A comparison of  $\Delta$(CN$-g^\prime$) vs.\  [Na/Fe] for RGB stars with spectroscopic
  abundances. The error bars for each point are shown in the bottom right of each panel. The red 
  lines show the best-fit line through the data using Equation~(\ref{eq:corr}) and the values 
  listed in Table~\ref{tab:correlation}.
  The data behind this figure is available in the article in machine readable format.
  }
\label{fig:corr}
\end{figure*}

To examine the sensitivity of the CN$-g^\prime$ color system to stellar population, we created
a fiducial red giant branch for each cluster.   We selected those RGB stars within the magnitude
range of the [Na/Fe] spectroscopy, and fit their locus with a third
order polynomial (see Figure~\ref{fig:rgb-cmd}) by minimizing the scatter in color using ordinary
least squares.  We then considered each star with a spectroscopic
[Na/Fe] determination, and correlated its distance from the fiducial line, $\Delta$(CN$-g^\prime$)
(measured color minus fitted color), with its spectroscopically determined value of [Na/Fe].
These comparisons are shown in Figure~\ref{fig:corr} and summarized in Table~\ref{tab:correlation}.

As expected, Figure~\ref{fig:corr} shows a correlation between CN$-g^\prime$ and
[Na/Fe] index:  higher metallicity stars have deeper CN absorption troughs, hence fainter
CN magnitudes and redder CN$-g^\prime$ colors.   For the three intermediate-metallicity, low-reddening
systems, the significance of this correlation is much greater than 99\% (according to the Pearson correlation
coefficient), and the slopes of the relation are similar.  In addition, the scatter of the 
data around the correlation is consistent with the reported uncertainties in [Na/Fe] 
and $\Delta$(CN$-g^\prime$). For M71, the correlation between color 
and [Na/Fe] abundance is less clear, possibly due to its higher overall metallicity and saturation 
of the CN-band \citep{suntzeff1981, langer1985}.  Our result for
M71 is consistent with that of \citet{smith2015b}, who measured a moderately strong correlation 
between spectroscopic CN-band strength and [Na/Fe] abundance in the cluster's RGB stars.  
For M5, we see a more robust correlation between CN strength and [Na/Fe] 
than was reported by \citet{smith2013}, likely due to our larger sample.

To quantitatively determine the dependence of CN$-g^\prime$ color on [Na/Fe], we followed the
recommendation of \citet{feigelson1992} and used the ordinary least squares (OLS) bisector method to
compute the slope of a $\Delta$(CN$-g^\prime$) vs. [Na/Fe] line.  This method consists of computing 
two OLS solutions,  in turn treating each variable as independent, and selecting the line that 
bisects these solutions.  The best-fit slope and its uncertainty is then found via a bootstrap analysis, 
which in our case, used 10,000 realizations. The corresponding intercept is found by minimizing the 
scatter in color. The best-fit line, shown in red in
Figure~\ref{fig:corr}, uses the slope and intercept given in Table~\ref{tab:correlation} and has the form
\begin{equation} \label{eq:corr}
{\rm [Na/Fe]} = ({\rm slope}) \times \Delta ({\rm CN}-g^\prime) + ({\rm intercept})
\end{equation}

\section{The Distribution of Abundances}

The width of the RGB (almost 0.6 mag) in the CN$-g^\prime$ color in each cluster shown 
in Figure~\ref{fig:rgb-cmd} is broader than that expected from photometric uncertainties (0.03 mag) and 
demonstrates that the CN$-g^\prime$ color system is sensitive to stellar abundance variations.
This finding was confirmed from photometry of artificial stars added along each 
cluster's fiducial giant branch in our $g^\prime$ and CN image frames.  
While there are other ways to gain this increased sensitivity, such 
as through the use of Str\"omgren photometry \citep[e.g.,][]{calamida2007}, Washington system 
photometry \citep[e.g.,][]{cummings2014}, broadband ultraviolet colors \citep{milone2017},
the CTIO Ca filter \citep{lim2015}, and measurements of calcium H \& K \citep[e.g.,][]{roh2011}, 
the CN measurements offer a fairly direct and efficient path to probe the range of 
light-element abundances in globular clusters.

%\floattable
\begin{deluxetable*}{ccccccc}
\tablewidth{0pt}
%\tabletypesize{\scriptsize} 
%\tablenum{3}
\tablecaption{Summary of the Correlation Between $\Delta$(CN$-g^\prime$) and [Na/Fe] \label{tab:correlation}}
\tablehead{
&\colhead{Number} &\colhead{Magnitude}   &\colhead{Correlation}  &\colhead{p-value} 
&\colhead{Slope} & \colhead{Intercept at} \\
\colhead{Cluster} &\colhead{of Stars}&\colhead{Limit}
&\colhead{Coefficient} &\colhead{two-tailed} &\colhead{[dex/mag]} & \colhead{$\Delta$(CN$-g^\prime) = 0$} }
\startdata 
 M3 & 35 &15.6 & 0.46 & 0.0053 & 2.29$\pm$0.50 & 0.13 \\
 M5 & 159 &16.7 & 0.63 & $8 \times 10^{-19}$ & 2.69$\pm$0.18 & 0.13 \\
 M13 & 70 &15.8 & 0.36 & 0.0023 & 2.93$\pm$0.91 & 0.26 \\
 M71 & 29 &17.7 & 0.39 & 0.0355 & 1.40$\pm$0.23 & 0.34 \\
% M3/M5/M13 &147 & -- & 0.41 & $2 \times 10^{-8}$ & 2.77$\pm$0.36 & 0.18 \\
\enddata
\end{deluxetable*}

\begin{figure*} %[h!] 
  \centering
  \subfloat[]{%
    \includegraphics[width=0.5\linewidth]{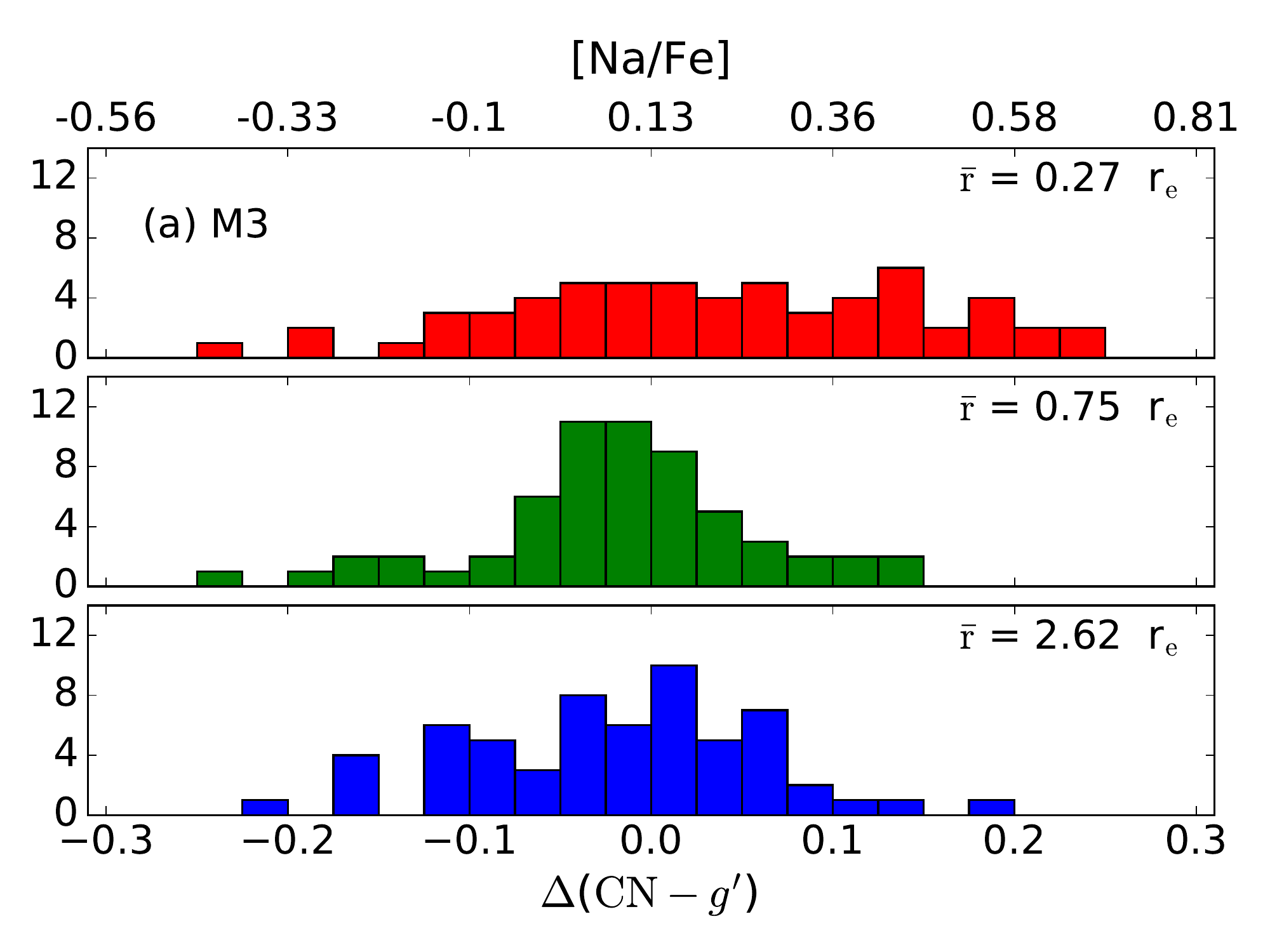}%
  }\hfill
  \subfloat[]{%
    \includegraphics[width=0.5\linewidth]{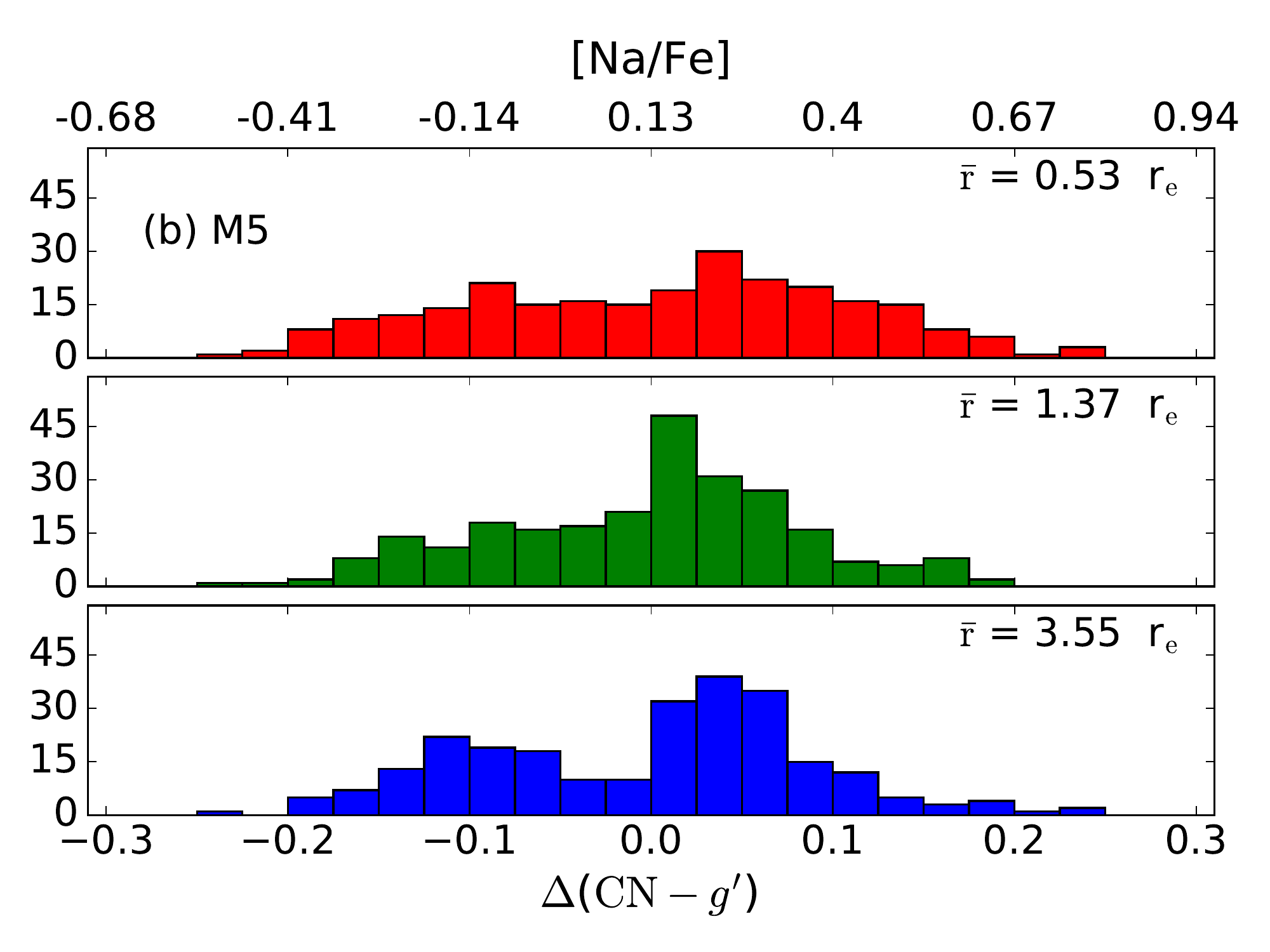}%
  }
  \\
  \subfloat[]{%
    \includegraphics[width=0.5\linewidth]{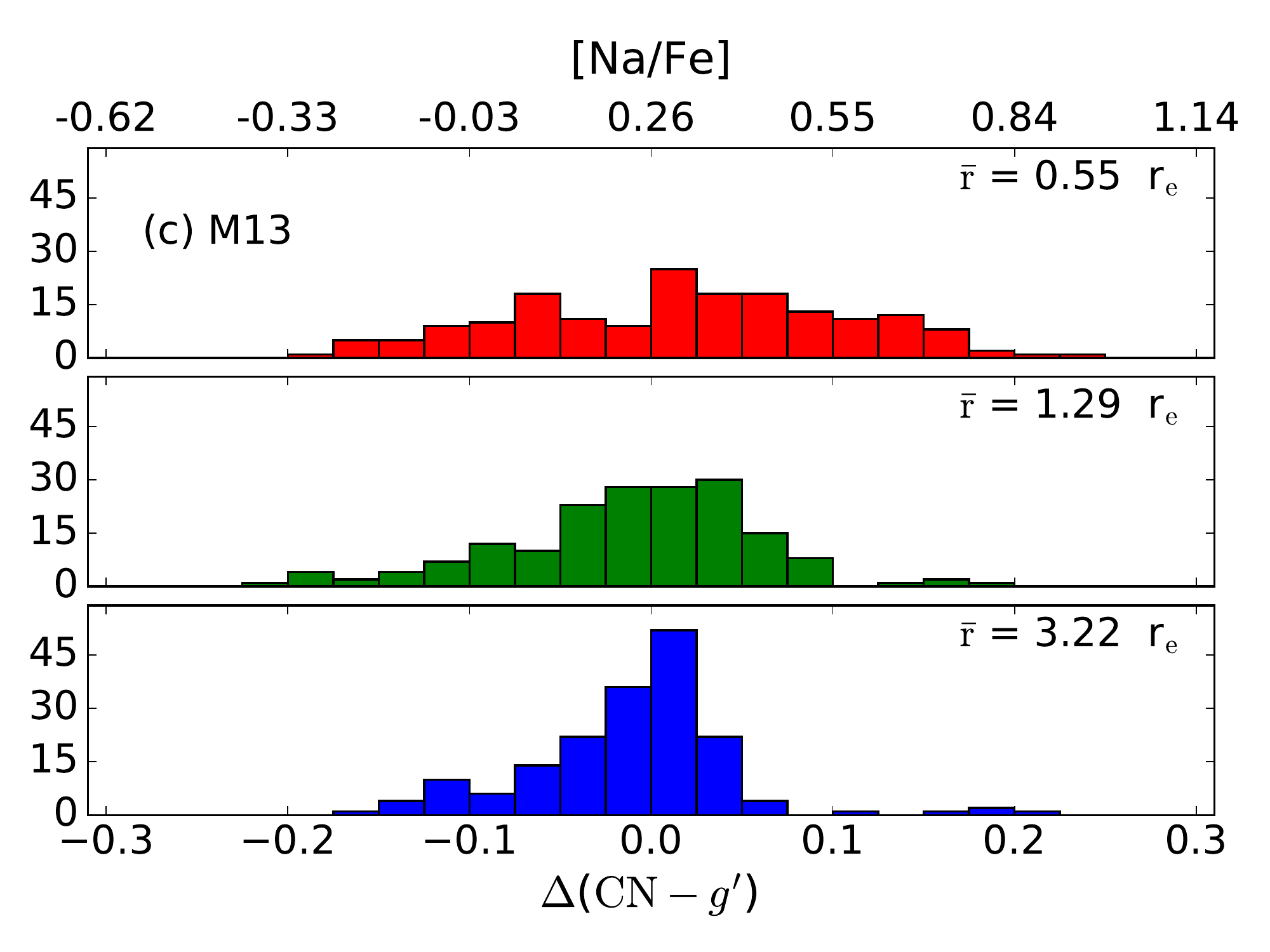}%
  }\hfill
  \subfloat[]{%
    \includegraphics[width=0.5\linewidth]{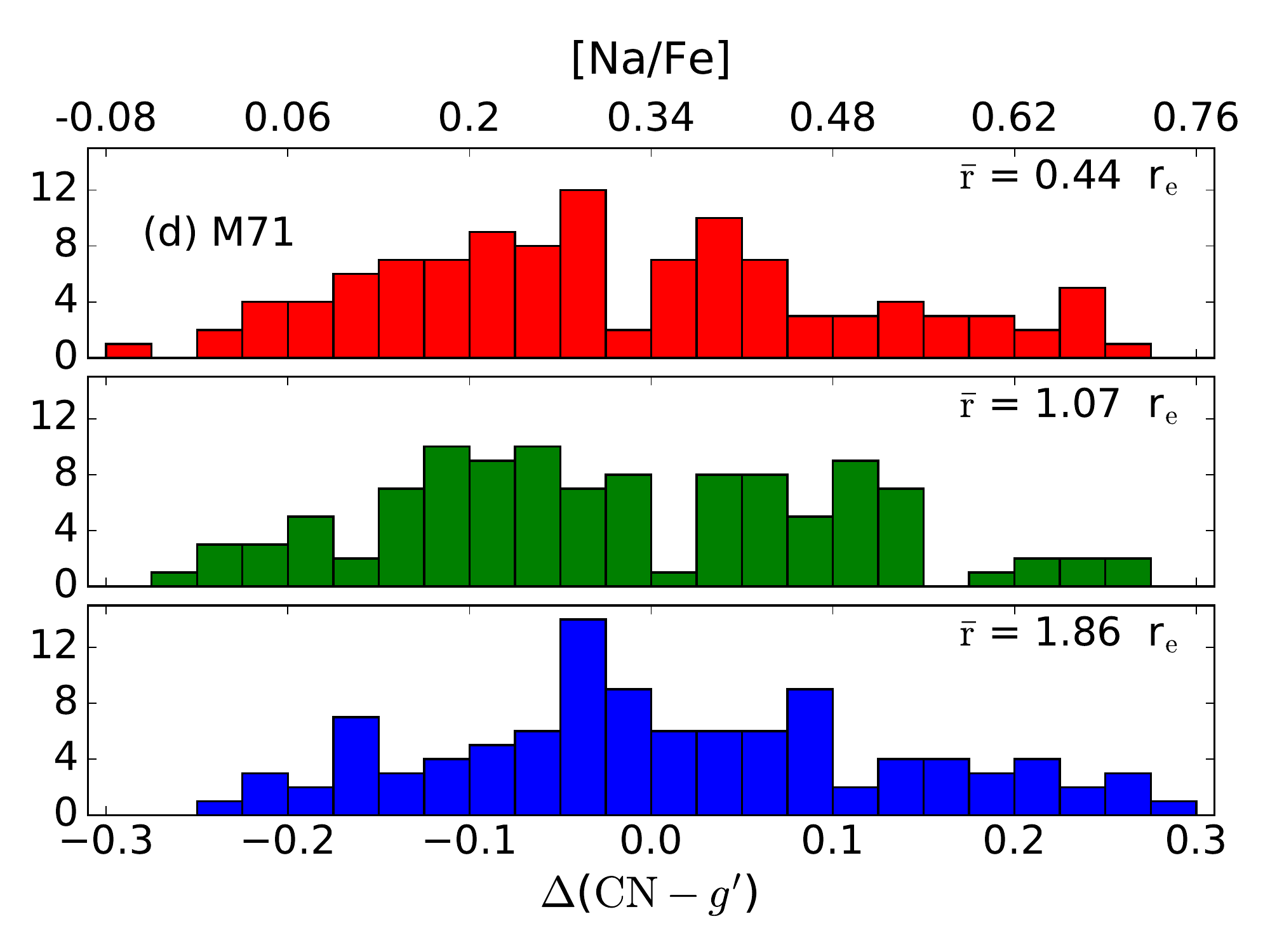}%
  }  
  \caption{The $\Delta$(CN$-g^\prime$) distributions for each cluster's RGB stars, divided into three
  radial bins, with equal numbers of stars in each radial bin, showing how CN strength varies with 
  radial distance from cluster center.  Both M3 and M13 show broader distributions of 
  $\Delta$(CN$-g^\prime$) color and an excess of CN-strong stars in the inner bin compared to the 
  outer two bins.  M5 is ambiguous, with a wider distribution of $\Delta$(CN$-g^\prime$) color in the
  inner bin but a similar average color compared to the outer two bins.  M71 shows no evidence for a 
  change in color, either in the average color or the dispersion in color, with cluster radius. 
  The [Na/Fe] scale along the top axis of each plot is calibrated for each cluster 
  using Equation~\ref{eq:corr} and the values in Table~\ref{tab:correlation}.  }
  \label{fig:radial}
\end{figure*}

The CN$-g^\prime$ photometry also allows us to study the radial dependence of abundance variations within
each cluster. Our results are illustrated in Figure~\ref{fig:radial}, where we have divided the RGB
stars of each system into three radial bins, each containing an equal number of stars.  The top 
axes of the plots translates these values into [Na/Fe] using Equation~\ref{eq:corr} and 
assumes that $\Delta$(CN$-g^\prime$) perfectly traces [Na/Fe] abundance. Even a
cursory examination of the data suggests that some degree of population segregation exists in most
clusters.  In particular, a Kolmogorov-Smirnov (K-S) comparison between the innermost and outermost
subsets demonstrates that the CN spread in the inner parts of M13, M3, and M5 is much wider than
that seen in the systems' outer regions. This result, which is significant at more than 95\% confidence 
(see Table~\ref{tab:ks}), agrees with the conclusions of previous studies 
\citep[e.g.,][]{carretta2010, johnson2012, cordero2014}, at least in the dynamically young clusters.  
There is also a notable deficit of stars with strong CN-bands in
the outer regions of M3 and M13, suggesting that the CN-enriched stars, which may be associated with
the cluster's second generation, are more centrally concentrated. 
The dependence is much weaker in M5, as the average $\Delta$(CN$-g^\prime$) values are roughly equal in 
the three radial bins. M71 displays no evidence for a population gradient, though given
the cluster's short dynamical timescale, this result is not surprising.  The lack of a population gradient in
M71 is also consistent with the [Na/Fe] analysis of  \citet{cordero2014}, who found a high degree of
spatial mixing between the primordial and enriched populations of the cluster.

\floattable
\begin{deluxetable}{cccc}
\tablewidth{0pt} 
%\tablenum{4}
\tablecaption{The radii dividing the RGB sample for each cluster into three bins containing an equal
number of stars \label{tab:bins}}
\tablehead{
\colhead{} & \colhead{Outer Radius of} & \colhead{Inner Radius of} & \colhead{Total stars in} \\
\colhead{Cluster} & \colhead{Inner Bin (r$_{e}$)} & \colhead{Outer Bin (r$_{e}$)} 
& \colhead{RGB sample}
}
\startdata
 M3  & 0.45 & 1.18 & 181 \\
 M5  & 0.87 & 2.04 & 762  \\
 M13 & 0.87 & 1.83 & 529 \\
 M71 & 0.75 & 1.38 & 331 \\
\enddata
%\tablecomments{See Table~\ref{tab:gc} for the half-light radius [${\rm r_e}$] of each cluster.}
\end{deluxetable}

\floattable
\begin{deluxetable}{cccc}
\tablewidth{0pt} 
%\tablenum{5}
\tablecaption{Results of a two-sample K-S test of the $\Delta$(CN$-g^\prime$) distributions between 
the inner and outer third of cluster stars\label{tab:ks}, and the fraction of second generation (SG)
stars in each radial bin}
\tablehead{
\colhead{} & \colhead{$p$-value} & \colhead{Mean(St.Dev) $\Delta$(CN$-g^\prime$)} 
& \colhead{SG stars (\%)} \\ 
\colhead{Cluster} & \colhead{(two-tailed)} & \colhead{Inner/Middle/Outer} & \colhead{Inner/Middle/Outer}
}
\startdata
 M3 &  0.0040 & 0.035(0.11) / -0.017(0.07) / -0.020(0.08) & 61 / 38 / 45 \\
 M5 &  0.044  & 0.004(0.11) / -0.002(0.08) / -0.002(0.09) & 55 / 57 / 58 \\
 M13 & $4.39 \times 10^{-11}$ & 0.017(0.09) / -0.010(0.07) / -0.011(0.06) & 62 / 48 / 47 \\
 M71 & 0.069 & -0.015(0.13) / -0.020(0.12) / 0.030(0.15) & 44 / 41 / 51 \\
\enddata
\end{deluxetable}

\begin{figure*} %[h!]
\centering
\noindent\includegraphics[width=0.8\linewidth]{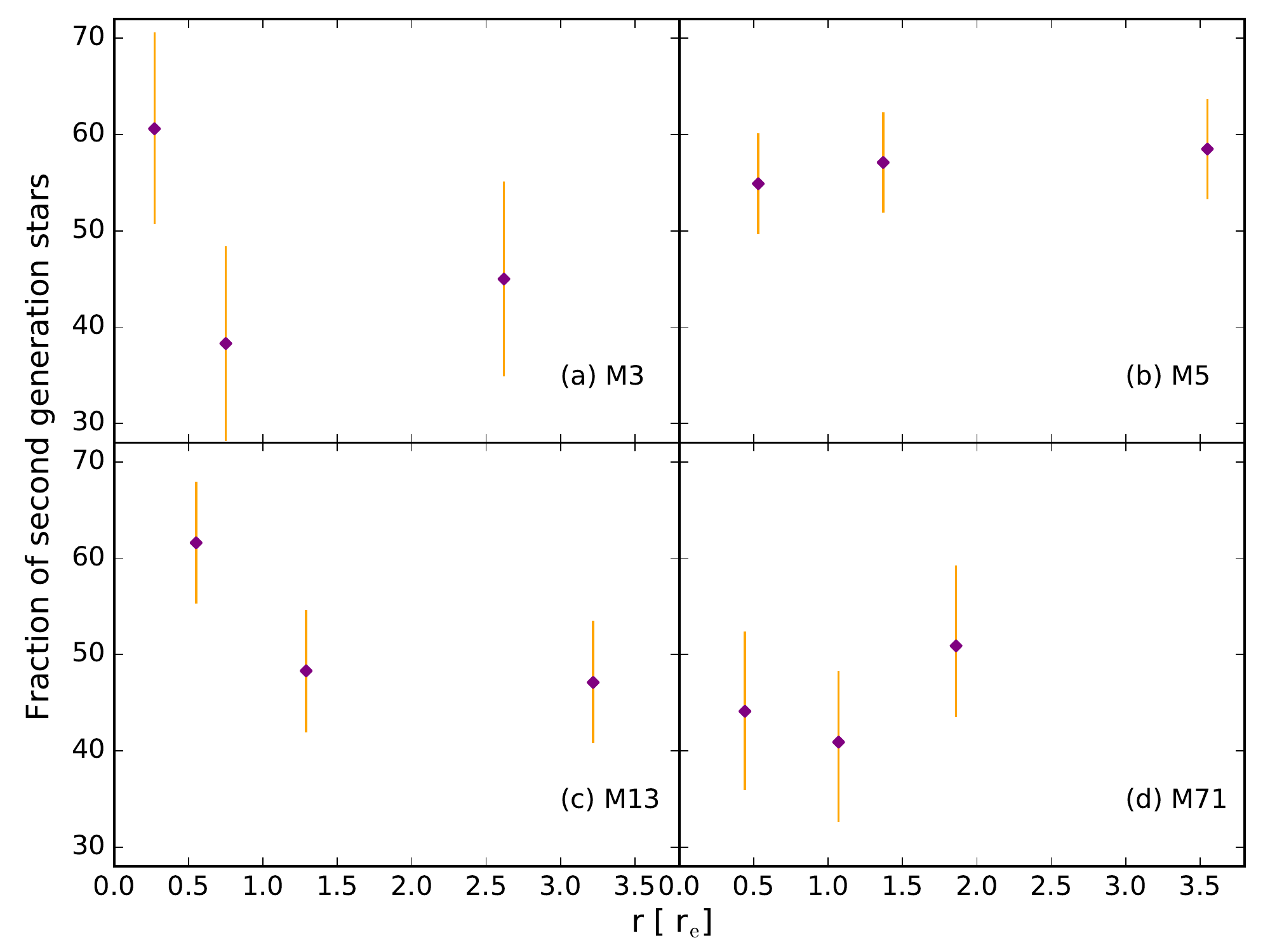}
\caption{The population ratio (second generation stars relative to total) for the three radial bins described
in Figure~\ref{fig:radial} and Table~\ref{tab:bins}. Second generation stars are identified as those 
lying redward of the fiducial RGB shown in Figure~\ref{fig:rgb-cmd} (i.e., stars with strong CN bands). 
The ratios are found via a bootstrap analysis using 10,000 samples, and the error bars reflect 90\% confidence
intervals. M3 and M13 show evidence for a change in the population ratio with radius, with second generation stars
more centrally concentrated than the primordial population.  The other two clusters show no such trend.}
\label{fig:pop-ratio}
\end{figure*}

\begin{figure*} %[h!]
\centering
\noindent\includegraphics[width=0.8\linewidth]{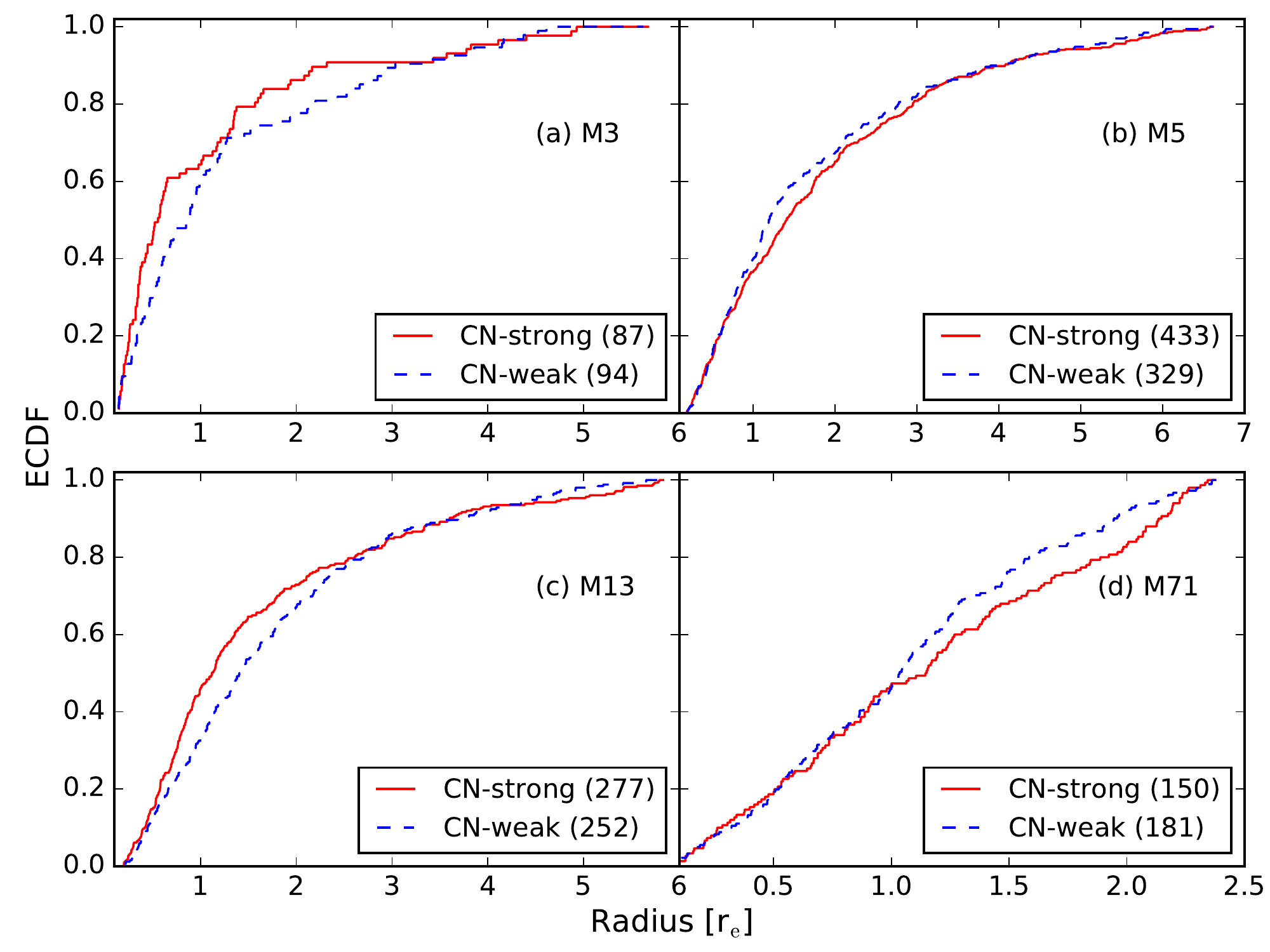}
\caption{The empirical cumulative distribution functions for CN-strong stars (red solid line) and 
CN-weak stars (blue dashed line) in each cluster, where the two groups are 
separated by the fiducial RGB curves shown in Figure~\ref{fig:rgb-cmd}. The number of stars 
in each sample are given in the plot legends. The CN-strong stars, which are likely second generation 
objects, appear to be more centrally concentrated in M3 and M13.  The CN-strong and CN-weak stars in M5
have statistically similar distributions. 
The dynamically relaxed cluster M71 shows no evidence for radial segregation.}
\label{fig:ecdf}
\end{figure*}

To further demonstrate population segregation in the clusters, we divided each cluster's
RGB stars into two samples: the CN-weak stars, which lie blueward of the mean RGB displayed
in Figure~\ref{fig:rgb-cmd} and are associated with the cluster's first generation, and 
CN-strong (second generation) stars, which fall redward of this fiducial curve. We examined the 
radial distribution of these samples by computing the population ratio
(the fraction of second generation stars) for each of the three radial bins 
described above and via the empirical cumulative distribution function (ECDF\null).  
These distributions are shown in Figure~\ref{fig:pop-ratio} and Figure~\ref{fig:ecdf}, respectively.  
The population ratio and corresponding uncertainty (90\% confidence intervals) are determined using a
bootstrap analysis (10,000 samples). 
The figures demonstrate that M3 and M13 have higher percentages of second generation, CN-strong stars in 
their inner regions than in their outer parts, confirming our measurement of the clusters' 
population gradient. There is no compelling evidence for this segregation in M5: the CN-strong 
and CN-weak stars have statistically similar distributions.
\citet{cordero2014} also found no radial dependence in the [Na/Fe] abundances in M5,
in contrast to the gradient detected by \citet{lardo2011}. In M71, the distributions of primordial 
and enriched stars appear to be identical, as is expected for a well-mixed cluster.

\section{Integrated Light}

The next step towards understanding the stellar populations of globular clusters is to
study these systems in other galaxies, where the conditions of formation may have been quite
different from that which occurred in the Milky Way.  This presents a problem, however, as at
distances larger than a few hundred kiloparsecs, spectroscopic measurements of individual stars
become prohibitively difficult.  However, the sensitivity of the CN-$g^\prime$ color index
gives us a tool with which to explore the chemical homogeneity of clusters, even at large distances.

\begin{figure}[h!]
\centering
\noindent\includegraphics[width=0.7\linewidth]{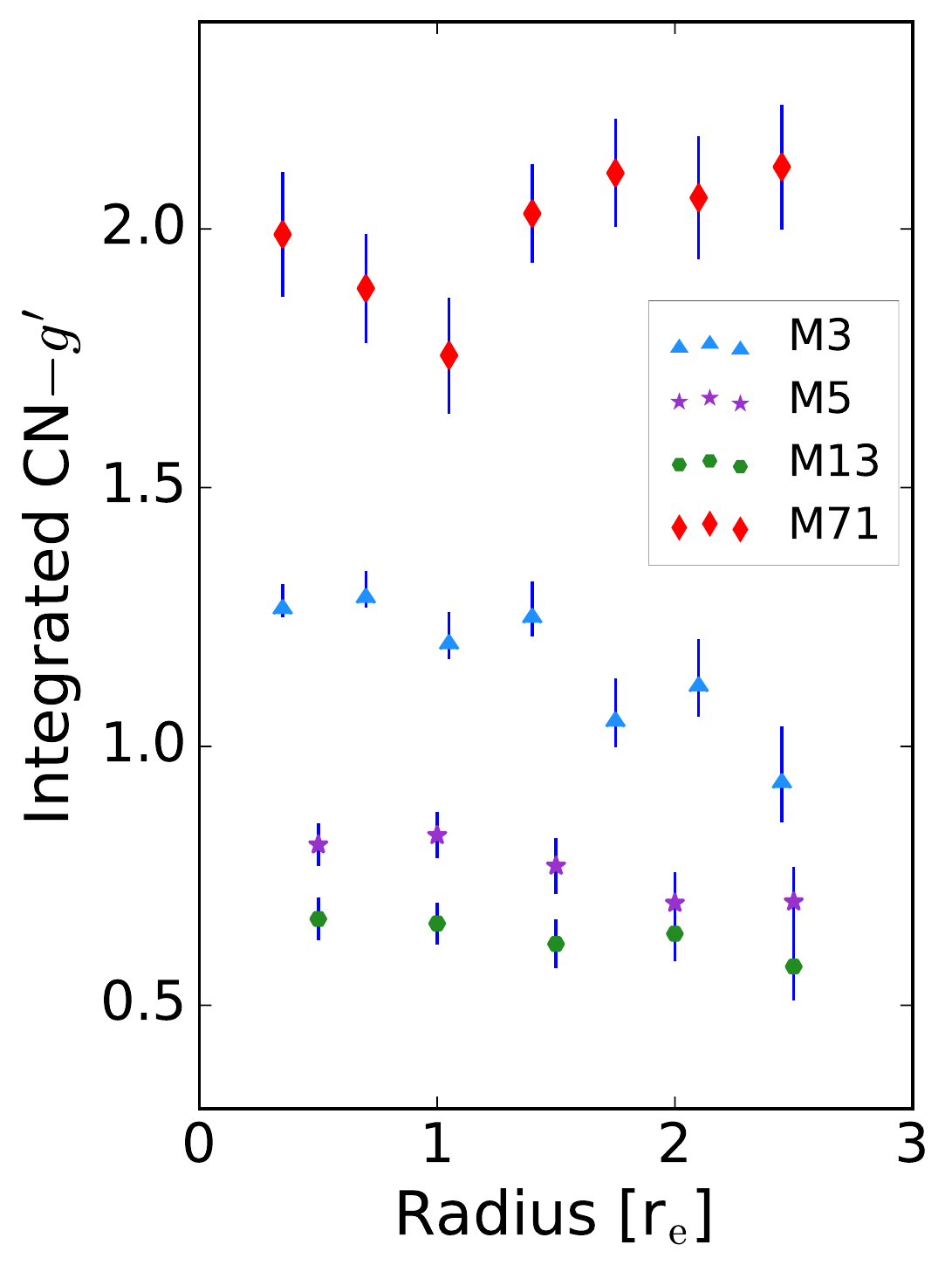}
\caption{Integrated CN$-g^\prime$ colors in different annuli are shown for each cluster.
  The CN zero point is based on the SDSS $u^\prime$ magnitude at the main-sequence turnoff.
  Errors on the points are estimated from the $g^\prime$
  luminosity contained in each region using Equation~\ref{eq:stoch}.
  Note that the photometric error bars on each color measurement are smaller than the points.
  In M3, the CN$-g^\prime$ color becomes bluer at larger radii, signifying fewer second generation 
  stars in the outer regions. (A greater number of CN-strong stars leads to redder colors, because the 
  enriched stars absorb more light in the blue CN-band).  Within our uncertainties, the integrated 
  colors of M5, M13 and M71 are independent of distance from cluster center. Our data are limited 
  to 2.5 effective  radii to avoid possible systematic errors associated with low surface brightness
  photometry.}
\label{fig:integrated-color}
\end{figure}

Figure~\ref{fig:integrated-color} mimics what might be recorded if the four globular clusters
were at the distance of M31 by displaying their integrated CN$-g^\prime$ colors
as a function of cluster radius.
From the figure, it is clear that the radial CN$-g^\prime$ color gradient, which is seen in
the individual RGB star abundances of the dynamically young system M3, is easily detected.
For M5 and M13, the integrated CN$-g^\prime$ colors remain constant to $\sim 1.5$ effective radii
but show evidence for a possible decline beyond this radius.
Conversely, M71, which is dynamically old and well mixed, shows no gradient either in the
individual RGB abundances or its integrated color.   This demonstrates the feasibility of
detecting multiple stellar populations within a globular cluster, even if that cluster is
too far away for its stars to be resolved. Note that we limit the radial extent of our integrated
color analysis to $\sim 2.5$ effective radii to avoid systematic offsets due to possible background
contamination at low surface brightness.

A reasonable concern with this integrated light approach is its sensitivity to stochastic processes
associated with the distribution of stars.   The integrated color of a stellar population 
depends on the number of bright objects present in the aperture.  For old systems, such as
globular clusters, this primarily means bright RGB and AGB stars, though under some conditions,
post-AGB stars and bright field objects may also be important.   In low luminosity systems (i.e., all GCs)
the number of these bright stars may be small, leading to color fluctuations, even in
homogeneous simple stellar populations.   Under the right circumstances, these stochastic effects
may mimic the signature of a change in stellar population.

To suppress the effect of stellar stochasticity in the gradient measurements of
Figure~\ref{fig:integrated-color}, the CN$-g^\prime$ color of each radial bin was derived from
the median of 19 different azimuthal slices within the cluster.   Unfortunately, when measuring
clusters outside of the Milky Way, it may not always be possible to use such fine subdivisions.
Thus, to understand quantitatively the effect that stochasticity has on the colors, we excluded
M71 from the analysis (due to the high density of field stars at its location) and used our data
on the remaining three intermediate-metallicity clusters to compare sets of color estimates
made at fixed radii but at different azimuthal position angles.  Specifically,  we divided each
radial annulus into several azimuthal bins and measured the color in each bin.  Because we expect
a cluster's stellar population to be constant within a given radial annulus, the changes in color
as a function of azimuth should provide a measure of the scatter introduced by stellar stochasticity.

\begin{figure*}
\centering
\noindent\includegraphics[width=0.6\linewidth]{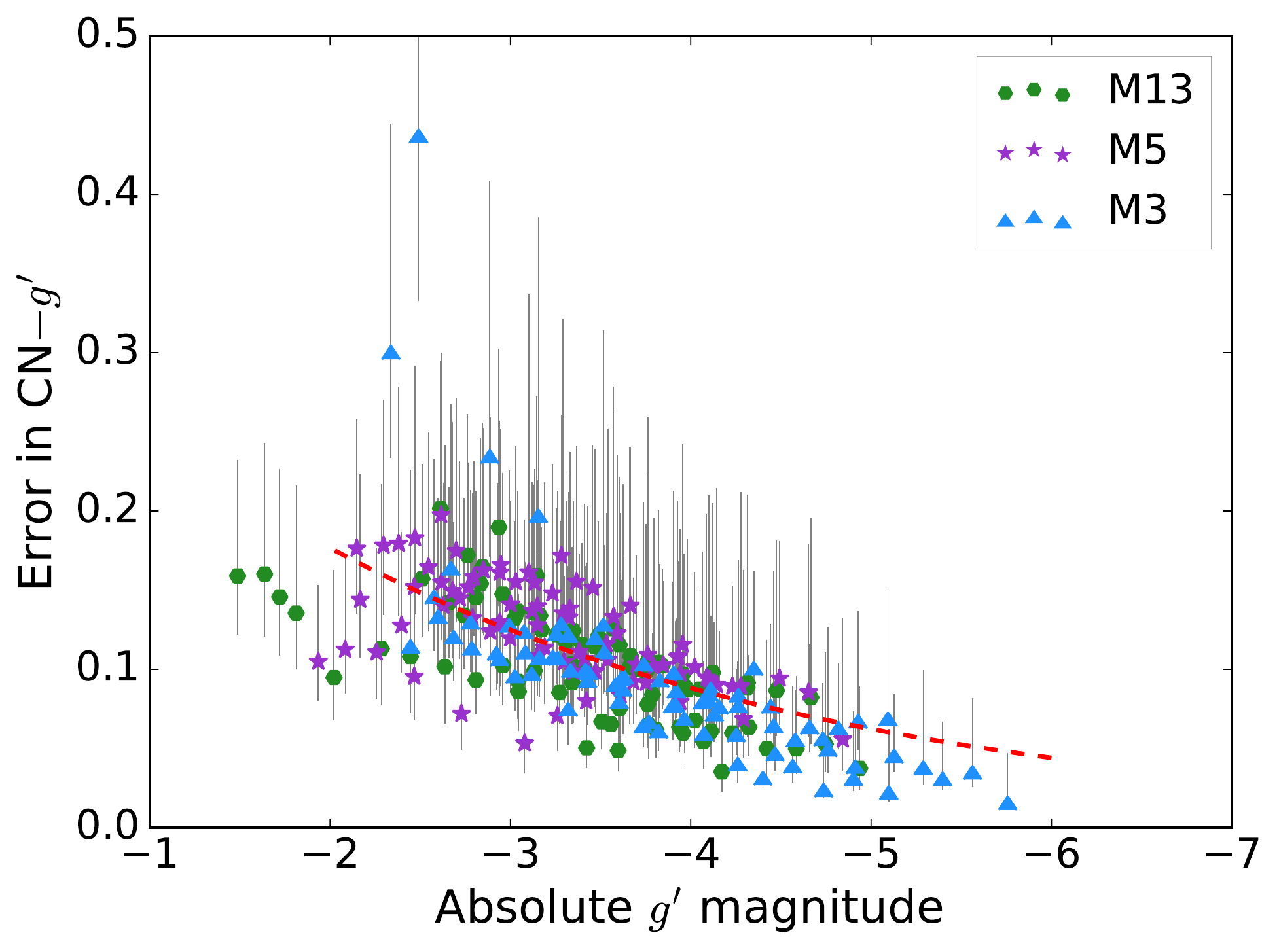}
\caption{The stochasticity in integrated CN$-g^\prime$ colors as a function of absolute $g^\prime$ 
magnitude. The CN zero point is based on the SDSS $u^\prime$ magnitude at the main-sequence turnoff.
The data were generated by dividing a given radial bin into several azimuthal 
subsections and measuring the color dispersion as a function of luminosity in the annulus. 
M71 has been omitted from the comparison, due to contamination from foreground field stars.  
Grey lines show the 90\% confidence intervals in the measured color dispersion. The red 
line gives the best fit in the form of an exponential decay.  This plot shows that for all systems 
with an integrated absolute $g^\prime$ magnitude brighter than $g^\prime \sim -5$, the uncertainty 
in the measurement of integrated CN$-g^\prime$ colors due to stellar stochasticity is quite small.}
\label{fig:integrated-error}
\end{figure*}

Figure~\ref{fig:integrated-error} displays the results of this analysis, along with the 90\% confidence
intervals on each dispersion measurement.  It is clear from the figure that for sample
regions of a cluster with an integrated absolute $g^\prime$ magnitude brighter than 
$g^\prime \sim -5$, the uncertainty introduced by measuring a finite number of stars is small.
If we fit an exponential decay model to the color error shown in Figure~\ref{fig:integrated-error},
then
\begin{equation} \label{eq:stoch}
\sigma({\rm CN}-g^\prime) \sim 0.35 \times 0.83^{-1.84 \times g^\prime}
\end{equation}
This is shown as the red dashed line in the plot.  Of course,
Figure~\ref{fig:integrated-error} represents a lower limit to the effect of stellar
stochasticity;  it is possible that when observing distant clusters, the survey will include some 
rare, bright sources that are not present in the finite populations of M3, M5, and M13.  Nevertheless,
the plot does demonstrate that under most circumstances,  the errors associated with
stellar evolution are minimal, and in our specific case, the color gradients displayed in 
Figure~\ref{fig:integrated-color} are real.

\section{Discussion}

The correlation of the $\Delta$(CN$-g^\prime$) color with published Na abundance, as shown 
in \S4, establishes the utility of the this color index to distinguish and 
track multiple populations in GCs.  In \S5, artificial star tests confirm that all four of the 
clusters observed show a wider distribution of $\Delta$(CN$-g^\prime$) color than would be expected 
from observational uncertainties, supporting the evidence of other photometric and spectroscopic
studies for the presence of multiple stellar populations in globular clusters.

The CN$-g^\prime$ color index provides us with an opportunity to efficiently characterize 
the stellar population gradients in GCs. As described in \S5,  we used individual stars to
examine histograms of $\Delta$(CN$-g^\prime$) in three radial bins and to compare the distribution 
of CN-weak and CN-strong stars in the inner and outer regions of the clusters. We also computed the 
population ratio (the fraction of second generation stars) for each of the three bins and compared 
the cumulative distribution functions for CN-strong and CN-weak stars as a function of distance 
from the clusters' centers.  
Finally, in \S6, we look for gradients in the CN$-g^\prime$
colors of our clusters using the systems' integrated light
that could be associated with changes in the fraction of CN-strong 
and CN-weak stars with distance from cluster center.  We now examine the results for each cluster in the 
context of dynamical evolution and compare our measurements with previous work.  
  
M3, with a present-day median relaxation time of 6.2 Gyr \citep{harris2010} and an age 
of 11.75 Gyr \citep{vandenberg2013}, is, dynamically speaking, the youngest of 
the globular clusters in our sample, as well as one of the youngest clusters of the Milky Way.
Histograms of the individual stars' $\Delta$(CN$-g^\prime$) index, divided equally into 
three radial bins (Figure~\ref{fig:radial}), show an excess of CN-strong stars in the inner bin, 
within $0.45 \, {\rm r_e}$ of the cluster center.

This result is corroborated by Figure~\ref{fig:pop-ratio}, 
which shows a second generation stellar population that is more centrally concentrated
than the primordial population. Likewise, the ECDF (Figure~\ref{fig:ecdf}) for M3 stars shows 
that CN-strong stars are more centrally concentrated than CN-weak stars. In integrated 
light (Figure~\ref{fig:integrated-color}), we find a clear gradient in the CN$-g^\prime$ color 
corresponding to a decrease in CN strength as one moves outwards from one to three effective radii.

Studies of stellar populations in M3 by \citet{johnson2005}, \citet{lardo2011}, 
and \citet{massari2016} provide evidence of significant differences in the central
concentration of its  stellar populations. As discussed in \S4, \citet{johnson2005} determined
Na abundances for several dozen stars in M3. Dividing the sample into half using their 
sodium abundances, we obtain a cumulative distribution function from the sodium abundance 
data that is qualitatively similar to the cumulative distribution function determined 
from our $\Delta$(CN$-g^\prime$) photometry; both show a clear central concentration of 
Na-rich stars compared to Na-poor stars.  We note that the stars analyzed by \citet{johnson2005} lie 
mostly outside of a radial distance of 0.5 effective radii from the center of M3, 
and their sample extends beyond 8 effective radii.

\citet{massari2016} used Str\"{o}mgren photometry, specifically the $c_{y}$ color, to separate 
the stellar populations of M3.  This index, defined by \citet{yong2008}, is sensitive 
to CN line strength, but not temperature.  M3's two populations are separated by about 0.05 mag 
in $c_{y}$ along the cluster's lower red giant branch.  Again, their cumulative distribution 
function for M3 is qualitatively similar to what we obtain using $\Delta$(CN$-g^\prime$) photometry.

\citet{lardo2011} examined the radial distributions of red giant stars in several globular clusters 
using ($u^\prime-g^\prime$) colors from SDSS (\citealt{abazajian2009} and references therein), as 
reanalyzed by \citet{an2008}.  Again, our cumulative distribution function for M3 is in 
qualitative agreement with that of \citet{lardo2011}, thereby confirming gradients in the population 
ratio in this cluster.

The presence of a population gradient in M3 is consistent with predictions.
\citet{vesperini2013} demonstrate that differences in the initial central concentrations of M3's stellar 
populations will survive at least until the present day.  For their least-centrally 
concentrated r2p5 model, with an initial ratio of the half-mass radius of first generation (FG) stars to 
the half-mass radius of second generation (SG) stars of $\mathrm{ R_{h,FG} / R_{h,SG} = 2.5 }$, the 
initial central concentrations remain near the original value for several relaxation times, 
and then begin to drop gradually over many relaxation times. 
For more highly concentrated initial conditions for second generation stars, the central concentration 
difference is erased more quickly.  In all of their simulations, however, the higher central concentration 
of second generation stars persists well beyond the dynamical age of M3.  As clusters age, 
second generation stars migrate outwards and the steepness of 
the population ratio gradient declines.  By two relaxation times, the population ratio is constant out to roughly 
0.5 half-mass radii, and declines beyond that point.  This general behavior is consistent with the decline in
M3's integrated CN$-g^\prime$ color seen in Figure~\ref{fig:integrated-color}.  It is important to note, 
however, that M3's initial central concentration of second generation stars is unknown.

M5's present-day median relaxation time of 2.6~Gyr \citep{harris2010} is less than a quarter
of its physical age (11.5~Gyr; \citealt{vandenberg2013}). It therefore represents an 
intermediate case, and evidence for its population gradient is ambiguous.  Histograms
of the individual stars' $\Delta$(CN$-g^\prime$) index (Figure~\ref{fig:radial}), 
divided equally into three radial bins, 
show a wider distribution of $\Delta$(CN$-g^\prime$) color in the
inner bin but a similar average color compared to the outer two bins.
The ECDFs (Figure~\ref{fig:ecdf}) for CN-weak and for 
CN-strong stars are similar, indicating little difference in the central concentration of its first and 
second generation stellar populations.  The population ratio plot given in Figure~\ref{fig:pop-ratio} also 
shows no change in the relative numbers of first and second generation stars with radius. The 
overall CN$-g^\prime$ color gradient for M5 (shown in Figure~\ref{fig:integrated-color}) is flat 
to $\sim 1.5$ half-light radii and then may decline slightly at larger radii.

Our ECDFs for M5 are consistent with an analysis of Na abundances by \citet{cordero2014}, including both her 
own measurements and those of \citet{ramirez2002}. As \citet{cordero2014} notes, the similarity in 
the cumulative distribution functions of first and second generation stars is discrepant with the analysis 
of ($u^\prime-g^\prime$) colors from SDSS photometry by \citet{lardo2011}, who find the second generation 
to be more centrally concentrated than the first generation population.  Cordero suggests that 
\citeauthor{lardo2011}'s sample better represents the stellar populations within one half-mass radius, 
while the spectroscopic data better sample the cluster at larger radii.

At the age and current relaxation time of M5, any initial difference between central concentrations 
of first and second generation stars will be significantly reduced.  If the difference was 
initially high, it would have largely been erased in the present-day cluster.  
If the difference was initially small, it could persist longer, 
but would be difficult to detect observationally.  After more than four dynamical time scales, the fraction
of second generation stars would be flat to beyond one half-light radius, and then decline slowly.  
This behavior is consistent with our observation of the change in CN$-g^\prime$ integrated color
out to 2.5 half-light radii (Figure~\ref{fig:integrated-color}).
  
The present-day median relaxation time for M13 is about 2 Gyr \citep{harris2010}, which is one-sixth of its 
12 Gyr lifetime \citep{vandenberg2013}.  As we found for M5, histograms of the $\Delta$(CN$-g^\prime$) color 
in radial bins (Figure~\ref{fig:radial}) show a broader distribution of stars in the central bin, 
but the mean color in the three bins is similar within our observational uncertainty.
The population ratios in Figure~\ref{fig:pop-ratio} show an excess of CN-rich
stars within ${\rm r_{e}} < 1$, and the ECDF curves for CN-rich and CN-poor stars in Figure~\ref{fig:ecdf} 
suggest that CN-rich stars are somewhat more centrally concentrated.
In integrated light (Figure~\ref{fig:integrated-color}), the radial dependence of the CN$-g^\prime$ 
color is flat within 1.5 half-light radii, and, as in M5, the CN$-g^\prime$ color declines 
slightly outside that radius.

\citet{johnson2012} examined the radial distributions of primordial, intermediate, and extreme stars in 
M13, finding relative populations of 15\%, 63\%, and 22\%, respectively in their sample.  They found that 
the extreme population is highly centrally concentrated, while the radial distributions of the primordial 
and intermediate populations are similar.  In order to compare more directly to our photometry, we have 
re-sorted the \citet{johnson2012} sample into two similarly sized subpopulations based simply on the 
stars' Na abundances.  We find little difference in the radial concentrations of the Na-rich and Na-poor 
stars when the sample is divided into equal halves.  
Moreover, the cumulative distribution functions from the Na abundance data are similar to what we find for 
our $\Delta$(CN$-g^\prime$) photometry when the stellar samples are divided into nearly equal halves.

While the clear central concentration of M13's extreme population persists \citep{johnson2012}, its 
intermediate and primordial populations are well mixed, exhibiting virtually no difference in central 
concentration. 
While the extreme population in M13 is highly centrally concentrated \citep{johnson2012}, 
primordial and intermediate stars still dominate within ${\rm r_{e} < 1}$ due to the relatively small 
fraction of extreme stars in the cluster ($\sim 15$\%).
As is the case for M5, either the initial difference in the central concentrations of M13's intermediate 
and primordial populations has been reduced by dynamical evolution within 1.5 half-light radii, or 
the central concentrations of these populations were similar to begin with.  Beyond 1.5 
half-light radii, the CN$-g^\prime$ color declines slightly, consistent with M13's dynamical age.
The lack of evidence for radial gradients in M5 and M13 is also consistent with the results 
from the study of the distribution of extreme blue and red horizontal branch 
stars in these clusters \citep{vanderbeke2015}.

M71, at an age of 11 Gyr \citep{vandenberg2013} and with a present-day dynamical relaxation time 
of 0.84 Gyr \citep{harris2010}, is the dynamically oldest cluster in our sample.  
Its central radial bin does not contain an excess of CN-strong stars (Figure~\ref{fig:radial}),
its ECDF shows no evidence for CN-strong stars being centrally concentrated (Figure~\ref{fig:ecdf}),
the population ratio gives no hint of segregation in the 
distribution of second generation stars (Figure~\ref{fig:pop-ratio}),
and there is no radial gradient in the cluster's integrated CN$-g^\prime$ 
color (Figure~\ref{fig:integrated-color}).
These results are consistent with the observations of \citet{cordero2015}, who found no 
statistical difference in the central concentrations of first and second generation stars in 
M71, as identified from their analysis of Na abundances in 75 cluster members.

At the advanced dynamical age of M71, the simulations of \citet{vesperini2013} suggest that the cluster 
should be well mixed, with no central concentration of its second generation stars, 
just as observed.  \citet{vesperini2013} 
also note that as a cluster approaches complete mixing, statistical fluctuations can add noise into the 
ECDF curves leading to small but detectable differences between populations, particularly in regions 
beyond 1-2 half-mass radii.  This effect could explain the separation of the CN-weak and CN-strong 
curves for M71 in Figure~\ref{fig:ecdf}.

Overall, the $\Delta$(CN$-g^\prime$) index provides an effective tool for exploring the radial 
distributions of stellar populations in globular clusters.  In each of the four clusters, M3, M5, M13, and M71, 
the index correlates with spectroscopically measured [Na/Fe] ratios and qualitatively reproduces published ECDFs 
for first and second generation stars.  Integrated light observations using a narrow-band filter at the 
CN-$\lambda 3883$ band allow the identification of stellar population gradients in clusters, with potential 
application to clusters where photometric measurements of individual stars may not be practical.  Combined with 
broadband photometric data from all-sky surveys such as SDSS \citep{an2008} and Pan-STARRS \citep{chambers2016}, 
and eventually from the Large Synoptic Survey Telescope, observations with a single narrow-band filter can 
identify and characterize stellar population gradients in large numbers of clusters.

\section{Summary}

A fuller understanding of the formation of globular clusters requires studying these systems in
galaxies with different formation histories.   Once outside the Galaxy, spectroscopic measurements of individual
stars are not possible --- one will have to probe for stellar population differences with integrated
light.  Our analysis of the globular clusters M3, M5, M13, and M71 demonstrates that CN$-g^\prime$
photometry is a viable way of making these measurements. There is a rich history of using the blue 
CN absorption feature to separate stellar populations, both spectroscopically and photometrically. Furthermore,
changes in the CN-band strengths can be observed at low resolution and even photometrically, while other
indicators like sodium or oxygen abundances require high-resolution spectra.

When we compare CN$-g^\prime$ colors with spectroscopically determined [Na/Fe] measurements for 
these four nearby globular clusters, we find clear evidence for a strong correlation between the
two quantities.  This relationship establishes that we can use the two-filter system to distinguish
stellar populations, and indeed, when we examine the systems' red 
giant branches, we find that all four have a spread in color that is many times broader than that 
associated with the photometric errors.  
Moreover, in the dynamically young, intermediate-metallicity cluster M3, we find evidence for a 
radial dependence in both the mean and the dispersion of the CN$-g^\prime$ color distribution,
in agreement with previous measurements \citep[e.g.,][]{carretta2010}. 
The situation is less clear for M5, for which the evidence for a gradient in CN strength as a 
function of radius is ambiguous.  For M13, the primordial and intermediate populations appear to 
be well mixed, although the extreme population is centrally concentrated.
No gradient is seen in M71, though given its old dynamical age, this result is unsurprising.

Finally, we examined the relationship between CN$-g^\prime$ measurements for individual RGB stars 
and similar measurements for integrated globular cluster light.   We showed that for M3,
integrated light measurements can detect the known CN$-g^\prime$ color gradient, and that the effect
of stellar stochasticity on the measurements is small in all systems, with an integrated 
absolute $g^\prime$ magnitude brighter than $g^\prime \sim -5$.
Future efforts will seek to apply this integrated light analysis to globular clusters in nearby
galaxies and allow us to better understand how subpopulations impact the formation and evolution of
these systems.  For example, the use of an integrated light approach should prove useful for probing
multiple populations in M31 globular clusters, especially clusters from the 
PAndAS catalog \citep{huxor2014} with half-light radii ($r_{e}$) exceeding 10\arcsec.  
In these systems, integrated CN$-g^\prime$ color gradients should allow us to detect population 
variations within clusters over a wide range of mass, dynamical age, and galactocentric radius.  
Population ratios can then be obtained from integrated color gradients in 
combination with detailed population synthesis. 

\acknowledgments
We are grateful to the anonymous referee for her/his critical and constructive comments, and for helping 
to improve the overall quality of this study. 
We thank Enrico Vesperini, Alex Hagen, Joanna Bridge, and Owen Boberg for their 
help, insight, and thoughtful comments as we prepared this manuscript.
The 0.9-m telescope is operated by WIYN Inc. on behalf of a Consortium of partner Universities
and Organizations (see www.noao.edu/0.9m for a list of the current partners). WIYN is a joint
partnership of the University of Wisconsin at Madison, Indiana University, Yale University,
and the National Optical Astronomical Observatory.  WPB \& CAP gratefully acknowledge support from
the Daniel Kirkwood endowment at Indiana University.  The Institute for Gravitation and the Cosmos is
supported by the Eberly College of Science and the Office of the Senior Vice President for Research at
the Pennsylvania State University.  This research has made use of the VizieR catalog access tool,
CDS, Strasbourg, France. The original description of the VizieR service was published in A\&AS 143, 23.
This research also made use of the USNOFS Image and Catalog Archive
operated by the United States Naval Observatory, Flagstaff Station
(http://www.nofs.navy.mil/data/fchpix/). This research has made use of NASA's Astrophysics Data
System Bibliographic Services.

\bibliography{bowman2017.ms.bib}

\end{document}